\documentclass[11pt,a4paper]{article}

\usepackage{jheppub}

\usepackage{amsmath}
\usepackage[normalem]{ulem}
\usepackage{bm}
\usepackage{dcolumn}
\usepackage{graphicx}

\newcommand{\beq}{\begin{eqnarray}}
\newcommand{\eeq}{\end{eqnarray}}

\begin{document}

\title{Topology in full QCD at high temperature:\\ a multicanonical approach}

\author[a,b]{Claudio Bonati}
\author[a,b]{Massimo D'Elia}
\author[c,d]{Guido Martinelli}
\author[b]{Francesco Negro}
\author[e]{Francesco Sanfilippo}
\author[a,b]{Antonino Todaro}

\affiliation[a]{Dipartimento di Fisica dell'Universit\`a di Pisa, Largo Pontecorvo 3, I-56127 Pisa, Italy}
\affiliation[b]{INFN sezione di Pisa, Largo Pontecorvo 3, I-56127 Pisa, Italy}
\affiliation[c]{Dipartimento di Fisica dell'Universit\`a di Roma ``La Sapienza'', Piazzale Aldo Moro 5, I-00185 Roma, Italy}
\affiliation[d]{INFN Sezione di Roma La Sapienza, Piazzale Aldo Moro 5, I-00185 Roma, Italy}
\affiliation[e]{INFN Sezione di Roma Tre, Via della Vasca Navale 84, I-00146 Roma, Italy}

\emailAdd{claudio.bonati@df.unipi.it}
\emailAdd{massimo.delia@unipi.it}
\emailAdd{guido.martinelli@roma1.infn.it}
\emailAdd{fnegro@pi.infn.it}
\emailAdd{sanfilippo@roma3.infn.it}
\emailAdd{antonino.todaro@pi.infn.it}

\abstract{
We investigate the topological properties of $N_f = 2+1$ 
QCD with physical quark masses, at temperatures
around 500 MeV. With the aim of obtaining 
a reliable sampling of topological modes in a regime 
where the fluctuations of the topological charge $Q$ 
are very rare, 
we adopt a multicanonical approach, adding a
bias potential to the action which enhances the probability 
of suppressed topological sectors.             
This method permits to gain up to three orders magnitude 
in computational power in the explored 
temperature regime. Results at different lattice
spacings and physical spatial volumes reveal no 
significant finite size effects and the presence, instead,
of large finite cut-off effects, with the topological susceptibility
which decreases by 3-4 orders of magnitude while moving from 
$a \simeq 0.06$ fm towards the continuum limit. 
The continuum extrapolation
is in agreeement with previous lattice determinations
with smaller uncertainties but obtained based on ansatzes justified by 
several theoretical assumptions. 
The parameter $b_2$, 
related to the fourth order coefficient in the 
Taylor expansion of the free energy density 
$f(\theta)$, has instead a smooth continuum extrapolation
which is in agreement with the dilute instanton gas 
approximation (DIGA); {moreover, a direct measurement of the 
relative weights of the different topological sectors
gives an even stronger support to the validity of DIGA.}
}


\maketitle

\section{Introduction}\label{intro}

The topological properties of QCD in the high temperature regime represent an
essential input to axion cosmology. The QCD axion, originally introduced to
explain the observed suppression of the topological CP-breaking
$\theta$-parameter in QCD~\cite{Peccei:1977hh, Peccei:1977ur, Wilczek:1977pj,
Weinberg:1977ma}, has {also been
recognized} as a possible candidate of dark matter. Through its coupling to the
topological charge density, the axion gets its effective (temperature
dependent) mass $m_a (T)$ from topological charge fluctuations, in particular
$m_a^2(T) = \chi(T)/f_a^2$, where $\chi(T)$ is the topological susceptibility
and $f_a$ is the effective axion coupling. Assuming axions as the
main source of dark matter, in particular through the so-called misalignment
mechanism~\cite{Preskill:1982cy, Abbott:1982af, Dine:1982ah}, a precise
knowledge of $\chi(T)$ for temperatures at or above the GeV scale 
{can give} access to the coupling constant $f_a$ and, in turn, to the
value of the axion mass today, which is a quantity relevant to present and
future experiments trying to detect it.
 
The analytic semiclassical predictions for $\chi(T)$ at asymptotically high $T$
come from the Dilute Instanton Gas Model~\cite{Gross:1980br, Schafer:1996wv,
Morris:1984zi, Ringwald:1999ze}, which is expected to become more and more
reliable as $T$ approaches the perturbative regime: it predicts a topological
susceptibility decaying as a power law in $T$, 
\begin{equation}
\chi(T) \propto T^{-D_2}
\label{eq:diga_chi}
\end{equation}
with $D_2 \simeq 8$ for three light flavors.  When trying to obtain
determinations which are reliable down to the GeV scale, non-perturbative
lattice computations are needed.  Lattice QCD simulations have provided
information on $\theta$-dependence below~\cite{Alles:1996nm, Alles:2000cg,
D'Elia:2003gr, Lucini:2004yh, DelDebbio:2004ns, Durr:2006ky, Giusti:2007tu,
Bazavov:2010xr, Luscher:2010ik, Panagopoulos:2011rb, Bazavov:2012xda,
Cichy:2013rra, Bruno:2014ova, Fukaya:2014zda, Cichy:2015jra, Ce:2015qha,
Bonati:2015sqt, Bonati:2016tvi} and across the deconfinement temperature 
$T_c$~\cite{Alles:1996nm, Alles:2000cg, Lucini:2004yh, Gattringer:2002mr,
DelDebbio:2004rw, Bonati:2013tt, Bonati:2015uga, Xiong:2015dya} since long, and
the interest related to axion physics has fostered renewed efforts from the
lattice community in the recent past~\cite{Berkowitz:2015aua, Kitano:2015fla,
Borsanyi:2015cka, Trunin:2015yda, axion_1, Petreczky:2016vrs, Frison:2016vuc,
Borsanyi:2016ksw, Burger:2018fvb}.  However, the investigation of the
topological properties of QCD in a regime of temperatures much larger than
$T_c$ has to face various non-trivial numerical challenges which can be
summarized as follows:

\begin{itemize}

\item[{\em i)}]
The topological susceptibility is usually determined starting from the
probability distribution $P(Q)$ of the topological charge $Q$, in particular
from its variance: $\chi(T) = \langle Q^2 \rangle/V_4$, where $V_4$ is the
space-time volume.  The suppression of $\chi(T)$ at large $T$ implies that, on
volumes reasonably achievable in lattice simulations, one may have $\langle Q^2
\rangle = V_4\, \chi(T) \ll 1$.  Since $Q$ is integer valued, that means that
configurations with $Q \neq 0$ become very rare, leading to the need of
unaffordably {large} statistics in order to achieve a correct
sampling of the topological charge distribution;

\item[{\em ii)}]
Topology in QCD with physical quark masses is strongly correlated with the
chiral properties of the theory, indeed $\theta$-dependence would be strictly
zero in the presence of massless quarks. That means that {the
explicit chiral symmetry breaking that is present in most fermion
discretizations} can lead to significant lattice artifacts, thus requiring very
small lattice spacings to achieve a reliable continuum extrapolation;

\item[{\em iii)}]
On the other hand, when trying to get closer to the continuum limit, a
different challenge emerges: because of the topological nature of the problem,
standard updating algorithms fail to correctly sample the distribution of $Q$
and get trapped in path integral sectors with fixed topology. This freezing of
topological charge leads to a severe critical slowing down of numerical
simulations~\cite{Alles:1996vn, DelDebbio:2002xa, DelDebbio:2004xh,
Schaefer:2010hu, Bonati:2017woi}; 

\item[{\em iv)}]
{Finally, since finite temperature 
numerical simulations are usually performed on lattices
$L_s^3 \times N_t$ with a fixed 
aspect ratio $L_s/N_t$, as $T = 1/(N_t a)$ becomes large also
the physical spatial volume $a^3 L_s^3$ becomes small, so that the 
possible presence of finite volume effects should be checked.
}

\end{itemize}

These are four different, partially independent and all potentially lethal
problems, which justify the discrepancies found in results from recent lattice
studies.  In particular, in Refs.~\cite{Trunin:2015yda, axion_1} the coefficient
$D_2$ of the power law suppression in Eq.~(\ref{eq:diga_chi}) was found to be
around 3, i.e.~much smaller than the DIGA prediction, in a range of
temperatures going up to around 600 MeV and exploring lattice spacings down to
around 0.06 fm. Within the misalignment mechanism for cosmological axion
production, a milder power law suppression implies a larger value of the
coupling constant $f_a$ in order to comply with the dark matter upper bound,
hence a smaller predicted value for the axion mass today. However, later
results found instead a substantial agreement with {the DIGA exponent,
even if larger by about one order of magnitude in absolute value}, 
in a range of
temperatures starting from very close to the pseudo-critical temperature $T_c$
and going up to a few GeVs~\cite{Petreczky:2016vrs, Borsanyi:2016ksw,
Burger:2018fvb}.

In particular, in Ref.~\cite{Borsanyi:2016ksw}, various strategies were devised
to circumvent the problems exposed above.  Problems {\em i)} and {\em iii)}
were bypassed at the same time, in the high-$T$ regime, by giving up sampling
the complete distribution $P(Q)$ of the topological charge from the path
integral, and determining instead the ratio of probabilities between fixed
topological sectors, in particular $P(\pm 1)/P(0)$, based on an integral method
which exploits numerical simulations in which the value of $Q$ is kept fixed on
purpose (see also Ref.~\cite{Frison:2016vuc}). This strategy is justified {provided
the main DIGA assumption is at work, i.e. that the distribution of instantons
and anti-instantons is that of non-interacting topological objects.} 
{Indeed, in this case, the distribution is Poissonian, so that one can learn everything about the 
distribution expected in the infinite volume limit even by exploring 
small volumes where the topological sectors $|Q| = 0,1$ are the 
only relevant ones; in absence of this strong assumption, one could 
be misled in deducing the infinite volume limit 
of $\langle Q^2 \rangle/V_4$ by a single measurement of 
$Z_1/Z_0$.}
On the
other hand this requires to know at least some properties of $P(Q)$, like for
instance the so-called $b_2$ coefficient, which is the fourth order coefficient
in the Taylor expansion of the free energy density as a function of 
$\theta$~\cite{Vicari:2008jw},
\begin{equation}
b_2 = - \frac{\langle Q^4 \rangle - 3 \langle Q^2 \rangle^2}
{12 \langle Q^2 \rangle}
\label{eq:b2}
\end{equation}
and approaches -1/12 when the DIGA sets in~\cite{Gross:1980br, Bonati:2013tt}.
{However, as we will discuss later on, $b_2$ is still 
not the end of the story because, even when DIGA fails,
it will turn out to be very close to -1/12 if the volume
is so small that just the $|Q| = 0,1$ sectors dominate
and moreover $Z_1/Z_0 \ll 1$. 
What one should really do is to check that the relative probability
ratios of different topological sectors (with $|Q| > 1$) follow the
volume
scaling predicted by DIGA. This requires, on the typical volumes
accessible with available computational power, to estimate events
(multiple instanton occurrences) which are order of magnitudes
smaller
than the already rare single instanton events.

In addition, 
in Ref.~\cite{Borsanyi:2016ksw}, trying to improve the convergence to
the continuum limit, a procedure was adopted which tries to correct for the
absence of exact zero modes in the lattice Dirac operator by suppressing gauge
configurations with $Q \neq 0$, using an ad hoc reweighting factor based on the
actual lowest eigenvalues of the adopted lattice Dirac operator.  
\\

The main purpose of this paper is to make progress towards an independent
determination of $\chi (T)$ in the continuum limit, trying to solve at least
problem {\em i)} from first principles and without any extra assumption.
To that purpose, we exploit a reweighting
technique which combines ideas typical of multicanonical
simulations~\cite{Berg:1992qua} and of metadynamics~\cite{LaioParr,
LaioGervasio, Laio:2015era}, and has already proved to be extremely efficient
in the toy model of the 1D quantum rotor~\cite{Bonati:2017woi}, where it
permits to correctly sample the distribution $P(Q)$ for $\langle Q^2 \rangle$
going down by several orders of magnitude, and more recently in pure gauge
theories~\cite{Jahn:2018dke}.

The main idea is to add a weight $\exp(-V(Q))$ to the path integral
distribution, where the potential $V(Q)$ is chosen so as to enhance the
probability of topological sectors which would otherwise be strongly
suppressed.  That is then corrected when computing averages over the
Monte-Carlo sample:
\begin{equation}
\langle O \rangle = \frac{ \langle O \exp(V(Q)) \rangle_V}
{ \langle \exp(V(Q)) \rangle_V}
\label{reweight1}
\end{equation}
where $O$ is a generic observable and $\langle \cdot \rangle_V$ stands for the
average taken according to the modified distribution.  Averages are left
unchanged, however fluctuations are modified leading to an improved
signal-to-noise ratio which, as we will show, permits to gain orders of
magnitude in terms of computational effort.

As already mentioned above, the idea is very similar in spirit to metadynamics,
where however the potential $V(Q)$ is modified dynamically during the
simulation, with the main purpose of enhancing tunneling between topological
sectors, thus defeating also the critical slowing down of topological modes.
Actually, our algorithmic framework was originally developed having a
metadynamical approach in mind; however, since freezing has not revealed to be
a severe problem in the explored range of lattice spacings and temperatures, we
have decided to opt for a static version of the potential $V(Q)$ which is
determined a priori. This framework is simpler because the problem comes back
to a standard equilibrium simulation and, on the other hand, the choice of the
potential does not reveal to be critical, apart from possible unwise 
choices which
however are easily avoided. 

The method is applied to numerical simulations of $N_f = 2+1$ QCD with physical
quark masses and for two different temperatures, $T \simeq 430$ and $T \simeq
570$ MeV, exploring lattice spacings down to $a \sim 0.03$ fm.  The improved
signal-to-noise ratio permits us to perform reliable infinite volume and
continuum limit extrapolations, which turn out to be in agreement with those of
Ref.~\cite{Borsanyi:2016ksw}.  Lattice artifacts reveal to be significant, so
that, despite the improved precision reached in simulations performed at finite
lattice spacing {and the absence of any ad hoc assumption
which could bias the result}, 
the final error for the continuum extrapolation is still
rather large.
Therefore, efforts to suppress lattice artifacts or to go to finer
lattice spacings by defeating the critical slowing down are still needed.

Another important achievement reached by our strategy is the 
possibility to obtain
a direct measurement of the relative weights of the 
different topological sectors, even if they 
are extremely small quantities on the accessible lattice volumes, 
thus giving a solid support to the validity of DIGA
in the explored temperature regime.

{Finally, we would like to stress that, due to computer limitations, 
we had to use staggered fermions, which require the rooting of the fermion determinant. 
In this paper we intended  to verify whether the reweighting procedure to enhance the lowest eigenvalues   and the limitation of the Montecarlo to the topological sectors with $Q=0,1$ would not bias the final result.  In the future further progress in our understanding of the complex dynamics of the topological charge  can   be obtained by adopting different lattice actions that do not require any rooting of the fermion determinant.}

{The paper is organized as follows. 
In Sec.~\ref{setup} we describe our numerical setup, regarding both 
the lattice discretization and the introduction and choice of the bias 
potential. In Sec.~\ref{results} we illustrate the improvement achieved
by our approach and present results for the topological susceptibility 
and the $b_2$ coefficient. Finally, in Sec.~\ref{discussion}, we discuss
perspectives for further improvement.}

\section{Numerical Setup}
\label{setup}

As in Ref.~\cite{axion_1}, we adopted a rooted stout staggered discretization
of $N_f = 2+1$ QCD, with a tree level improved Symanzik action~\cite{weisz,
curci} for the pure gauge sector.  The standard, finite temperature partition
function reads 
\begin{equation}
\mathcal{Z} = \int \!\mathcal{D}U \,e^{-\mathcal{S}_{Y\!M}} \!\!\!\!\prod_{f=u,\,d,\,s} \!\!\! 
\det{\left({M^{f}_{\textnormal{st}}[U]}\right)^{1/4}}
,  \label{partfunc}
\end{equation}
where
\begin{equation}
\mathcal{S}_{Y\!M} = - \frac{\beta}{3}\sum_{i, \mu \neq \nu} \left( \frac{5}{6}
W^{1\!\times \! 1}_{i;\,\mu\nu} -
\frac{1}{12} W^{1\!\times \! 2}_{i;\,\mu\nu} \right) \label{tlsyact} 
\end{equation}
and $W^{n\times m}_{i;\ \mu\nu}$ denotes the trace of the $n\times m$ Wilson
loop in the $\mu,\nu$ plane and starting at site $i$, constructed in terms of
the original gauge links of the theory, which are the integration variables in
Eq.~(\ref{partfunc}).  The staggered fermion matrix
$M^{f}_{\textnormal{st}}[U]$ reads instead
\begin{equation}
(M^f_{\mathrm{st}})_{i,\,j}= am_f \delta_{i,\,j}+ \sum_{\nu=1}^{4}\frac{\eta_{i;\,\nu}}{2}
\left[U^{(2)}_{i;\,\nu}\delta_{i,j-\hat{\nu}} -U^{(2)\dagger}_{i-\hat\nu;\,\nu}
\delta_{i,j+\hat\nu}  \right] \
\label{fermmatrix}
\end{equation}
and is constructed in terms of the modified link variables $U^{(2)}_{i,\mu}$,
which are obtained after two levels of the stout-smearing procedure introduced
in Ref.~\cite{Morningstar:2003gk}, with isotropic smearing parameter
$\rho=0.15$. 

Stout smearing is a convenient smoothing procedure which shares with other
smoothing techniques the benefits of suppressing lattice artifacts, in
particular by reducing the taste violations present in the staggered
discretization.  However, at a variance with other smoothing techniques, stout
smeared links are analytic functions of the original gauge link
variables~\cite{Morningstar:2003gk}, so that a standard Rational Hybrid
Monte-Carlo (RHMC) algorithm~\cite{Clark:2004cp, Clark:2006fx, Clark:2006wp}
can be easily applied.

The bare parameters, $\beta$, $m_s$ and $m_l\equiv m_u=m_d$, have been chosen
so as to move on a line of constant physics {(LCP)}, with a physical value of
the pseudo-Goldstone pion mass, $m_{\pi}\approx 135\,\mathrm{MeV}$ and of the
strange-to-light mass ratio, $m_s/m_l \simeq 28.15$. {The LCP parameters have
been fixed} according to the determinations reported in Refs.~\cite{physline1,
physline2, physline3} or to a cubic spline interpolation of them. In
Table~\ref{tab:bareparam} we provide a summary of our simulation points,
including the lattice sizes $L_s^3 \times N_t$ and the temperatures $T = 1/(a
N_t)$.

\begin{table}
\centering
\begin{tabular}{|c|c|c|c|c|}
  \hline
  \rule{0mm}{3.2mm} $\beta$ & $a$ [fm] & $a m_s$  & $L_s^3 \times N_t$ 
& T [MeV] \\ \hline
\rule{0mm}{3.2mm}4.140&0.0572&\(2.24\times 10^{-2}\)&\(24^3,32^3,40^3\times 8\) & 430\\ \hline
\rule{0mm}{3.2mm}4.280&0.0458&\(1.81\times 10^{-2}\)&\(24^3,32^3,40^3\times 10\) & 430\\ \hline
\rule{0mm}{3.2mm}4.385&0.0381&\(1.53\times 10^{-2}\)&\(28^3,36^3,48^3\times 12\) & 430\\ \hline
\rule{0mm}{3.2mm}4.496&0.0327&\(1.29\times 10^{-2}\)&\(32^3,40^3,48^3\times 14\) & 430\\ \hline
\rule{0mm}{3.2mm}4.592&0.0286&\(1.09\times 10^{-2}\)&\(36^3,48^3,64^3\times 16\) & 430\\ \hline
\rule{0mm}{3.2mm}4.140&0.0572&\(2.24\times 10^{-2}\)&\(24^3 \times 6\) & 570\\ \hline
\rule{0mm}{3.2mm}4.316&0.0429&\(1.71\times 10^{-2}\)&\(32^3 \times 8\) & 570\\ \hline
\rule{0mm}{3.2mm}4.459&0.0343&\(1.37\times 10^{-2}\)&\(40^3 \times 10\) & 570\\ \hline
\rule{0mm}{3.2mm}4.592&0.0286&\(1.09\times 10^{-2}\)&\(48^3\times 12\) & 570\\ \hline
\end{tabular}
\caption{Simulation 
parameters used in this work. Bare parameters
have been fixed according to Refs.~\cite{physline1,
    physline2} or spline interpolation of data thereof. The
  systematic uncertainty on the lattice spacing determination is
  $2-3\%$ and the light quark mass is fixed by 
  $m_s/m_l=28.15$.} 
\label{tab:bareparam}
\end{table}

\subsection{The multicanonical algorithm}

As already outlined above, in order to efficiently sample the topological
charge distribution $P(Q)$ in a regime where fluctuations away from $Q = 0$
become very rare, we will modify the Monte-Carlo weight of gauge
configurations, adding to the distribution appearing in Eq.~(\ref{partfunc}) a
$Q$-dependent bias, i.e.~sampling according to 
\begin{equation}
{\cal P[U]} \mathcal{D}U \propto
\mathcal{D}U 
\,e^{-\mathcal{S}_{Y\!M}} \!\!\!\!\!\!
\prod_{f=u,\,d,\,s} \!\!\!\! 
\det{\left({M^{f}_{\textnormal{st}}[U]}\right)^{1/4}}\, 
\! \! e^{-V(Q_{mc}[U])}  
\label{modified_weight}
\end{equation}
where $V(Q_{mc})$ is the bias potential and $Q_{mc}$ is a proper discretization
of the topological charge which will be in general different from the
observable $Q$ measured to determine the cumulants of the topological charge
distribution.

Monte-Carlo averages obtained in this way, denoted by $\langle \cdot
\rangle_V$, can then be combined to estimate standard averages based on
Eq.~(\ref{partfunc}), using the reweighting formula
\begin{equation}
\langle O \rangle = \frac{ \left\langle O\, e^{V(Q_{mc})} \right\rangle_V}
{ \left\langle e^{V(Q_{mc})} \right\rangle_V}
\label{reweight2}
\end{equation}
In principle, the choice of $V$ is not critical, since
after reweighting one recovers the original averages anyway.  However, on one
hand we would like to construct $V$ so as to enhance topological sectors which
would be otherwise strongly suppressed, in order to improve the statistical
accuracy in the determination of $\langle Q^2 \rangle$. On the other hand, we
would like to avoid a typical problem of reweighting, i.e.~a possible bad
overlap between the distribution in Eq.~(\ref{partfunc}) and that in
Eq.~(\ref{modified_weight}), which could result in a failure to sample
important configurations.  Both issues will be discussed in more detail in the
next subsection, where we show how $V$ has been chosen in practice. 

A different, important issue regards the  discretized topological charge
$Q_{mc}[U]$ entering the potential.  Several different lattice implementations,
gluonic or fermionic, are available in principle. However, the need for an easy
implementation of the potential into the RHMC Molecular Dynamics equations
constrains the choice.  Indeed, the introduction of $V\left(Q_{mc}\right)$
induces a new force term $F$ related to the value of $Q_{mc}$. By using the
chain rule one obtains
\begin{equation}
F_\mu(i) \equiv -\frac{\partial V}{\partial U_\mu(i)}=-\frac{\partial V}{\partial Q_{mc}}\,\frac{\partial Q_{mc}}{\partial U_\mu(i)} \, .
\label{chain_1}
\end{equation}
Therefore, the calculation of the new force term related to a given gauge link
proceeds through the calculation of the scalar coefficient ${\partial
V}/{\partial Q_{mc}}$, and the calculation of the derivative of {$Q_{mc}$} with respect to the given gauge link. 

In order to keep the calculation of the derivative simple, an easy choice
{would be} to consider the field-theoretical clover-based
definition of the topological charge~\cite{DiVecchia:1981qi, DiVecchia:1981hh}:
\begin{equation}
Q_{ft} = \sum_i q_{ft} (i) \ ; \quad 
q_{ft} (i) = -\frac{1}{2^9 \pi^2} 
\sum_{\mu\nu\rho\sigma = \pm 1}^{\pm 4} 
{\tilde{\epsilon}}_{\mu\nu\rho\sigma} \hbox{Tr} \left( 
\Pi_{\mu\nu}(i) \Pi_{\rho\sigma}(i) \right) \; ,
\end{equation}
where $\Pi_{\mu\nu}$ is the plaquette operator,
$\tilde{\epsilon}_{\mu\nu\rho\sigma}$ is the Levi-Civita tensor for
positive entries and is fixed by antisymmetry and 
${\tilde{\epsilon}}_{\mu\nu\rho\sigma} =
-{\tilde{\epsilon}}_{(-\mu)\nu\rho\sigma}$ otherwise.
With such a definition, the calculation of the force term driving
the dynamics amounts to a simple modification of the usual Wilson gauge action
force, in which the staples are decorated {by clover insertions}, 
{just as} in pure gauge simulations 
implementing an imaginary $\theta$ term~\cite{Panagopoulos:2011rb}.

However, despite the easy implementation, such a
definition is not convenient for the purpose of this study, 
in the naive way above.
Indeed, it must be remembered that
$Q_{ft}$, like any field-theoretic definition, is related  to the actual
topological background $Q$, configuration by configuration, through the
relation
\begin{equation}
Q_{ft}=Z Q + \eta,
\end{equation}
where $Z$ is a multiplicative renormalization 
constant~\cite{Campostrini:1988cy} and $\eta$ is a noise term which is practically
independent of $Q$ and has zero expectation value. 
It is well known that the renormalization constant $Z$ is typically
quite small, while the variance of $\eta$ is quite large, with $\langle \eta^2
\rangle \gg Z^2 \langle Q^2 \rangle$, so that $Q_{ft}$ has a small correlation
with the actual topological background $Q$, i.e.~
\begin{equation}
\frac{\langle Q Q_{ft} \rangle}
{\sqrt{\langle Q^2 \rangle \langle Q_{ft}^2 \rangle}}
\simeq \sqrt{\frac{ Z^2 \langle Q^2 \rangle}
{Z^2 \langle Q^2 \rangle + \langle \eta^2 \rangle}}
\, \ll 1 \, . 
\label{eq:correlation}
\end{equation}
While this is not harmful to the reweighting procedure, which is exact anyway
as long as one uses the same charge in the RHMC and in Eq.~(\ref{reweight1}),
it makes it useless, since the bias potential will not be able to
{increase the variance of} the sampled distribution
of topological charge $Q$, because of
the loose correlation with it\footnote{The situation when using an imaginary
$\theta$ term is more favourable in this respect, since the aim is to change
the average of the distribution and not its variance, in which case even
$Q_{ft}$ can be effective.}.

In order to circumvent this problem, an improved definition of $Q_{mc}$ is
needed. This can be easily achieved by standard ``smoothing'' techniques,
i.e.~computing $Q_{ft}$ on smoothed gauge fields in place of the original ones.
{This} has the effect of moving $Z$ closer to 1, and of reducing the
fluctuations of the noise $\eta$, thus leading to a substantial improvement for
the correlation with $Q$.  Many methods have been proposed so far to smooth
gauge configurations, including the gradient flow~\cite{Luscher:2009eq,
Luscher:2010iy}, cooling~\cite{Berg:1981nw, Iwasaki:1983bv, Itoh:1984pr,
Teper:1985rb, Ilgenfritz:1985dz}, and several kinds of smearing techniques, all
being equivalent to each other~\cite{Bonati:2014tqa, Cichy:2014qta,
Namekawa:2015wua, Alexandrou:2015yba, Alexandrou:2017hqw, Berg:2016wfw}. Given
the need of integrating the equations of motion induced by $V(Q_{mc})$, the
most natural choice is again to make use of the same stout smearing procedure
adopted to define the fermion matrix, which makes $Q_{mc}$ a differentiable
function of the original gauge links. 

In particular, assuming to compute $Q_{ft}$ on {$n$}
times stout-smeared links $U^{(n)}_\mu(i)$, the computation of the
``topological force'' related to the bias potential proceeds through 
\begin{equation}
\frac{\partial Q_{mc}}{\partial U_\mu(i)} =
\frac{\partial Q_{ft}[U^{(n)}]}{\partial U^{(n)}_\nu(j)} \,
\frac{\partial U^{(n)}_\nu (j)}{\partial U^{(n-1)}_\rho (k)} \,
\dots \, \frac{\partial U^{(1)}_\sigma (l)}{\partial U_\mu (i)} \, . 
\label{chain_2}
\end{equation}
The actual implementation of this chain relation amounts to the same procedure
used to compute the derivative of the pseudo-fermion contribution to the force,
{described} in Ref.~\cite{Morningstar:2003gk} and
{already} adopted in our {fermion} discretization. For small 
{values of the stouting parameter}
$\rho_{st}$, $n_{st}$ stout-smearing steps are equivalent to a gradient flow
time $\tau = n_{st}\, \rho_{st}$~\cite{Alexandrou:2017hqw}, hence to a number
of cooling steps $n_{cool} = 3\, \tau = 3\, n_{st}\, \rho_{st}$~\cite{Bonati:2014tqa}.

One would like to have $n_{st}\, \rho_{st}$ large enough to enhance $Z$ and
suppress $\eta$, so as to have a reasonable correlation between $Q_{mc}$ and
$Q$. However, using a large number of stout smearing steps implies a large
numerical overhead in the implementation of the molecular dynamics equations
while, on the other hand, 
{stout smearing does not act as a smoothing if $\rho_{st}$ is too large}~\cite{Morningstar:2003gk}.
As a
matter of fact, we found as a good compromise to fix $\rho_{st} = 0.1$ and to
choose $n_{st}$ in the range from 10 (for the finest lattice spacing) to 20
(for the coarsest), which corresponds to a number of equivalent cooling steps
going from 3 to 6. For this choice, the total numerical overhead due to the
introduction of the bias potential goes from a minimum of 30\% to a maximum of
60\%, measured with the respect to the computational time needed in absence of
the bias potential.  
Notice that, despite the moderate overhead, implementation of
the bias potential in full QCD, through the RHMC equations, does
not require a change of simulation paradigm as it is needed in
the pure gauge case~\cite{Jahn:2018dke}.

Finally, the determination of the topological charge $Q$ used for measurements
was based on a standard cooling procedure, adopting in particular $n_{cool} =
80$, then rounding it to the closest integer value, in particular following the
procedure {originally proposed in Ref.~\cite{DelDebbio:2002xa}} 
{(see also Ref.~\cite{Bonati:2015sqt} for more details).}
{Such a definition is based on the expectation that the physical
topological content of gauge configurations becomes stable
under action minimization when the continuum limit is approached,
while ultraviolet (UV) fluctuations responsible for renormalizations
are removed, 
and has been proved to yield results equivalent to the gradient
flow}~\cite{Bonati:2014tqa, Alexandrou:2015yba}. 
{Results obtained in this work 
have been checked to
be indeed stable, 
within statistical errors, in a range of cooling steps going from 40
to 120; however, in order to be sure to correctly 
include possible systematics of the method, 
the observed small variations in this range has been added as a
systematic error in all cases.}

\subsection{Practical implementation and choice of the potential}

{In this section, as} an illustrative example, we 
{discuss the way in which we fixed the
form of the biasing potential for the case of } the $32^3 \times 8$ lattice at
$\beta = 4.140$ (see Table~\ref{tab:bareparam}), corresponding to $T \simeq
430$ MeV. 

In Fig.~\ref{fig:mchistory_4.1403208standard} we report the Monte-Carlo
history of the topological charge, from which it is already clear that only
rare fluctuations to just the $Q = \pm 1$ topological sectors take place, this
is also clear looking at Fig.~\ref{fig:histostout_4.1403208standard}, where we
report the probability distribution of the topological charge measured after 20
steps of stout smearing. Indeed, we obtain $\langle Q^2 \rangle \simeq 0.01$,
meaning that the sectors with non-zero topological background are suppressed by
about 2 orders of magnitude with respect to the $Q = 0$ sector.  In particular,
taking also autocorrelations into account, we obtain $a^4\, \chi = \langle Q^2
\rangle / V_4 = (4.1 \pm 1.6) \times 10^{-8}$, which is compatible with the value
obtained at the same temperature and lattice spacing in Ref.~\cite{axion_1}.

\begin{figure}[t!]
\centering
\includegraphics[width=0.65\columnwidth, clip]{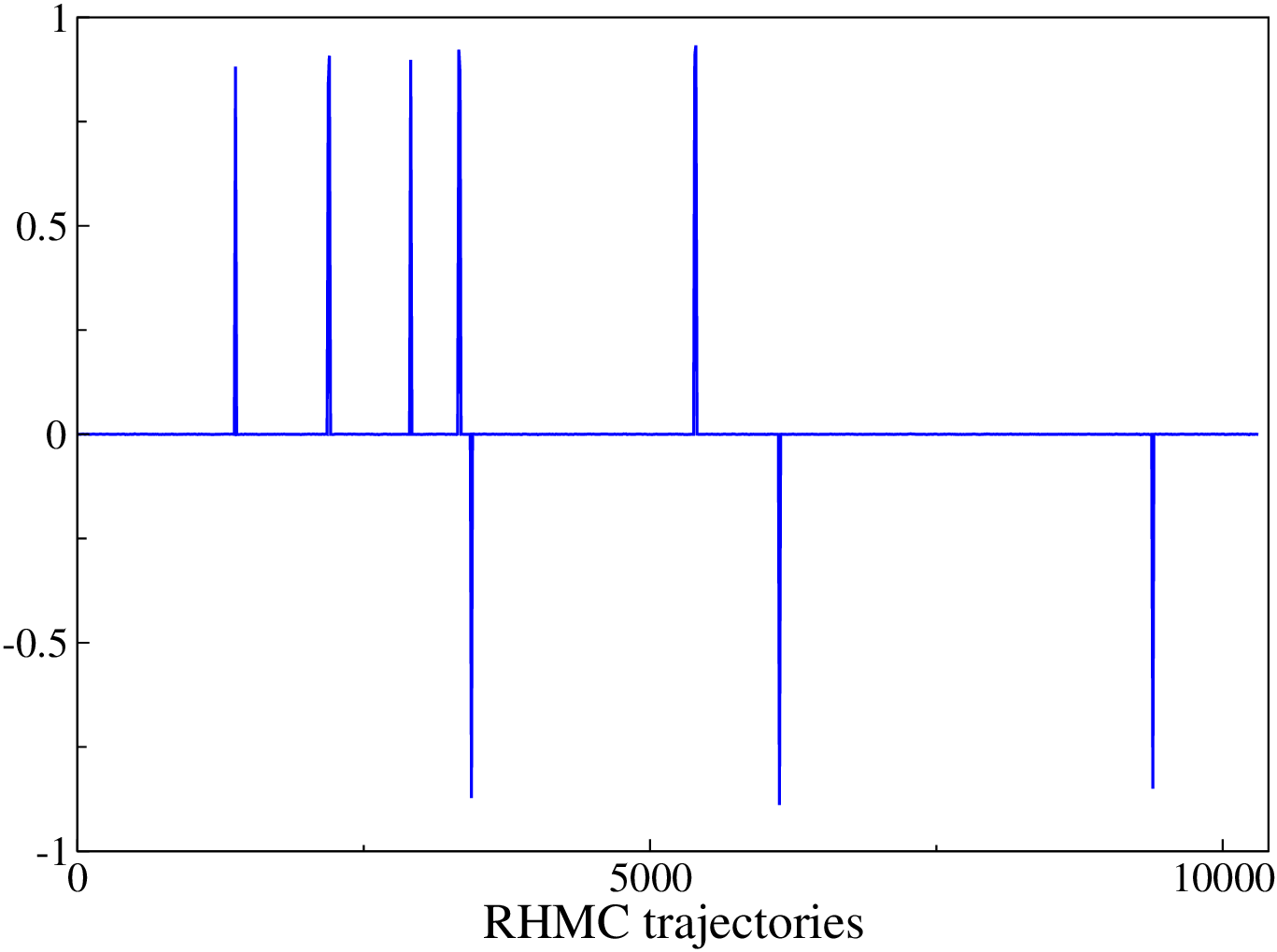}
\caption{Monte-Carlo history of the topological charge measured after 80
cooling steps and adopting the standard updating algorithm, i.e.~with no bias
potential, for a $32^3 \times 8$ lattice at $\beta = 4.14$.  Measurements are
taken every 10 RHMC trajectories, and $\langle Q^2 \rangle \sim
0.01$, meaning that the sectors with non-zero topological background are
suppressed by about 2 orders of magnitude.  } 
\label{fig:mchistory_4.1403208standard}
\end{figure}

\begin{figure}[h!]
\centering
\includegraphics*[width=0.65\columnwidth, clip]{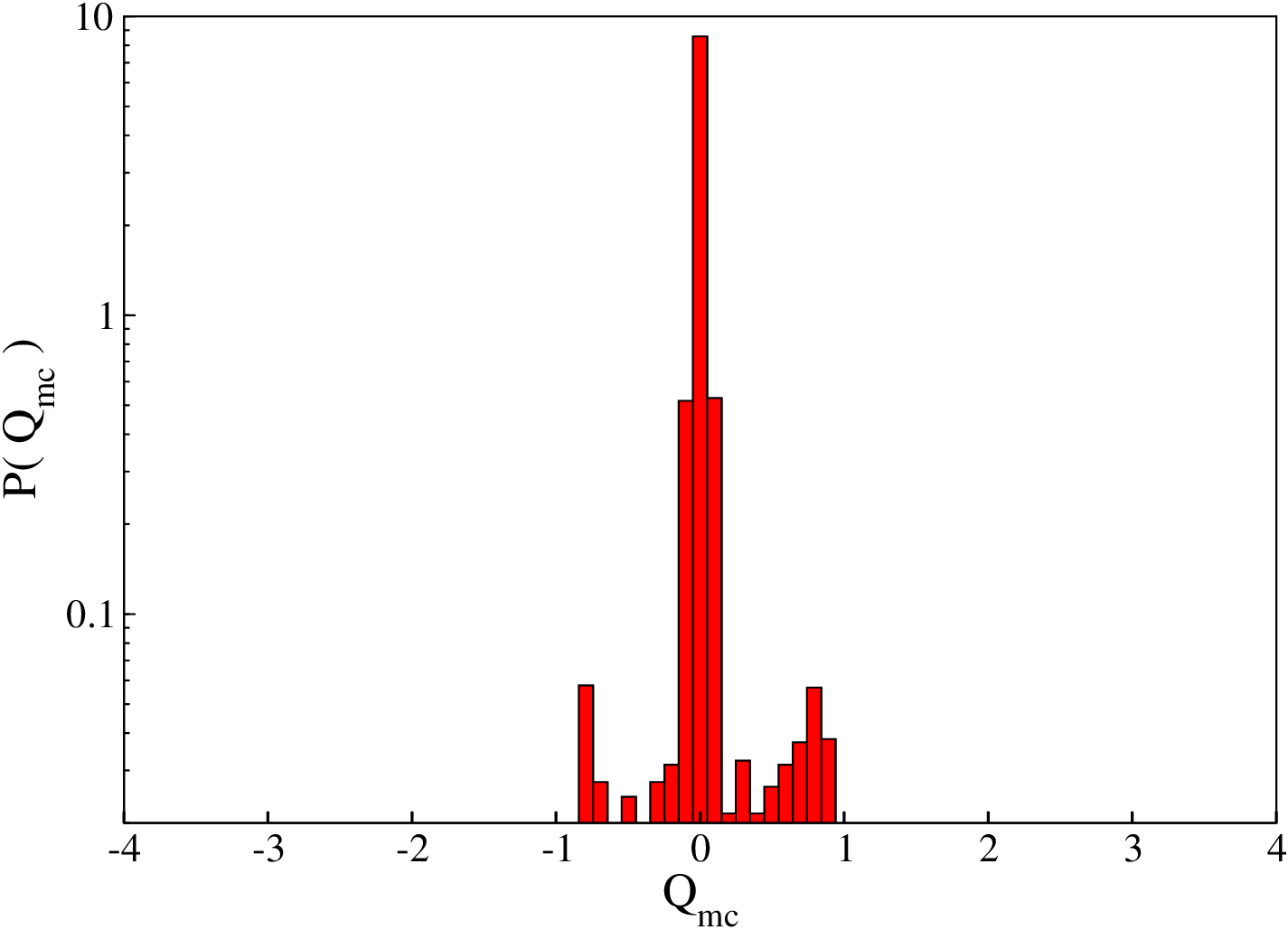}
\caption{Probability distribution of the topological charge $Q_{mc}$ measured
after 20 stout smearing steps with $\rho_{st} = 0.1$, for the same run
parameters relative to Fig.~\ref{fig:mchistory_4.1403208standard}. The
probability distribution is plotted in logarithmic scale, to better visualize
the peaks corresponding to a non-zero topological background, which are
suppressed by about 2 orders of magnitude.
}
\label{fig:histostout_4.1403208standard}
\end{figure}

In principle, the optimal choice for $V(Q_{mc})$ would be the one which makes
the biased topological charge distribution flat, i.e.~$V(Q_{mc})$ should be
taken equal to minus the logarithm of the probability distribution for $Q_{mc}$
determined in absence of the potential: that would enhance the probability of
configurations at the border between different topological sector, thus
defeating critical slowing down at the same time. However, this would require a
precise a priori knowledge of $P(Q_{mc})$: were that available, the problem
would have already been solved.  On the other hand, we are not looking  for the
optimal choice but just for a substantial improvement, so we will explore some
suitable smooth potentials.

A first possibility is to choose a quadratic potential,
\begin{equation}
V(Q_{mc}) = - a_q\, Q_{mc}^2 \, 
\label{potquad}
\end{equation}
since, at least for large enough volumes, the probability distribution of
$Q_{mc}$ is expected to become Gaussian-like.  Willing to enhance the $Q = \pm
1$ sectors by a factor $O(100)$, as it seems necessary looking at
Fig.~\ref{fig:histostout_4.1403208standard}, one should choose $a_q \sim
\log(100) \simeq 4.6$; however, even smaller magnitudes of $a_q$ may lead to
dangerous instabilities.  Indeed, as an example, consider a potential like that
in Eq.~(\ref{potquad}), with $a_q = 3.25$, which is shown in
Fig.~\ref{fig:potenziali}, where the bias towards larger
topological charges is stopped at a threshold charge $Q^{max} = 3$, fixing
$V(Q_{mc}) = V(Q^{max})$ for $|Q_{mc}| \geq Q^{max}$.  The Monte-Carlo history of
the topological charge obtained with this bias potential is shown in
Fig.~\ref{fig:storiasbagliata}: $Q_{mc}$ gets trapped around $Q^{max}$ and as a
result also $Q$ assumes only positive values. That means that the bias is too
strong towards large values of $|Q_{mc}|$, so that the biased system develops a
sort of spontaneous breaking of CP symmetry: that clearly affects the validity
of the reweighting procedure, since topological sectors which are relevant to
the original path integral now are badly sampled\footnote{Indeed, trying to
apply Eq.~(\ref{reweight2}) to this sample, we obtain a determination of
$\langle Q^2 \rangle/V_4$ which is about one order of magnitude larger than the
one found in the standard simulation, with an error which is even two orders of
magnitudes larger than that, meaning that it is completely out of control.}.

\begin{figure}[t!]
\centering
\includegraphics*[width=0.65\columnwidth, clip]{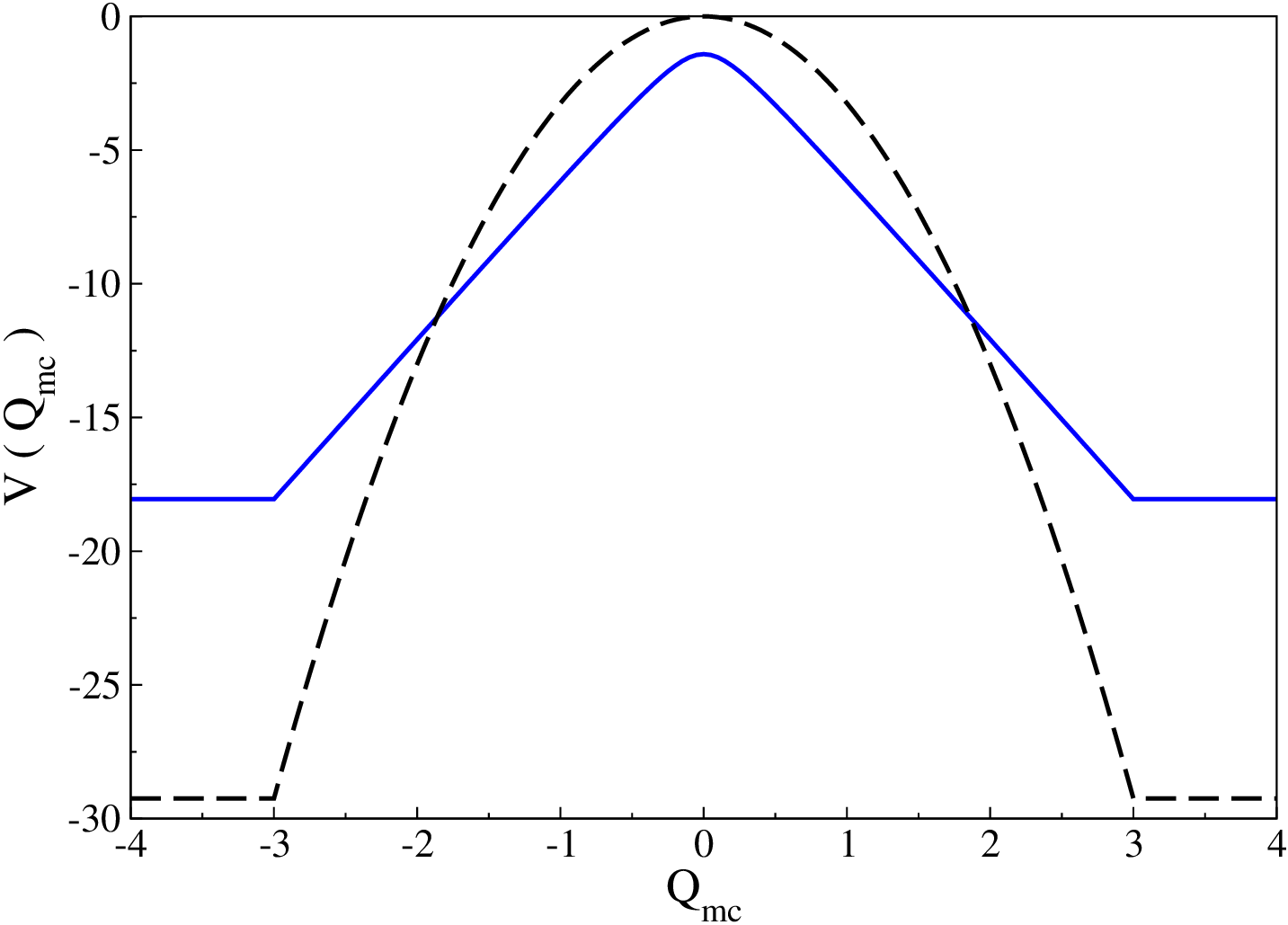}
\caption{Bias potentials adopted in the illustrative example discussed in the
text. The continuous line refers to the potential in Eq.~(\ref{potquad}), with
$a_q = 3.25$ and a cut-off at $Q^{max} = 3$. The dashed line refers to the
potential in Eq.~(\ref{potrel}), with $B = 6$, $C = 2$ and a cut-off at
$Q^{max} = 3$.
}
\label{fig:potenziali}
\end{figure}

It is of course possible to choose lower values of $a_q$ for which this problem
does not appear, however, in order to maintain a good enhancement of low $Q
\neq 0$ sectors, we have explored a different class of potentials with a less
steep behavior at large $Q$, in particular
\begin{equation}
V(Q_{mc}) = - \sqrt{(B\, Q_{mc})^2 + C} \, .
\label{potrel}
\end{equation}
In Fig.~\ref{fig:potenziali} we show one example of such potential, which we
have found as a good compromise after a few trial runs, for which $B = 6$ and
$C = 2$: also in this case the bias is cut at $Q^{max} = 3$. The
corresponding Monte-Carlo history is shown in Fig.~\ref{fig:storiagiusta}: the
topological charge now fluctuates evenly around $Q = 0$, and sectors which
are important in the original partition function are well sampled, however also
contributions from $Q \neq 0$ are explored frequently, so that it is possible
to obtain a more accurate determination of their contribution to $\langle Q^2
\rangle$ after using Eq.~(\ref{reweight2}). This is also visible from the
probability distribution obtained for $Q_{mc}$ during the biased run
{shown in Fig.~\ref{fig:histostout_4.1403208stout}}. 

\begin{figure}[t!]
\centering
\includegraphics*[width=0.65\columnwidth, clip]{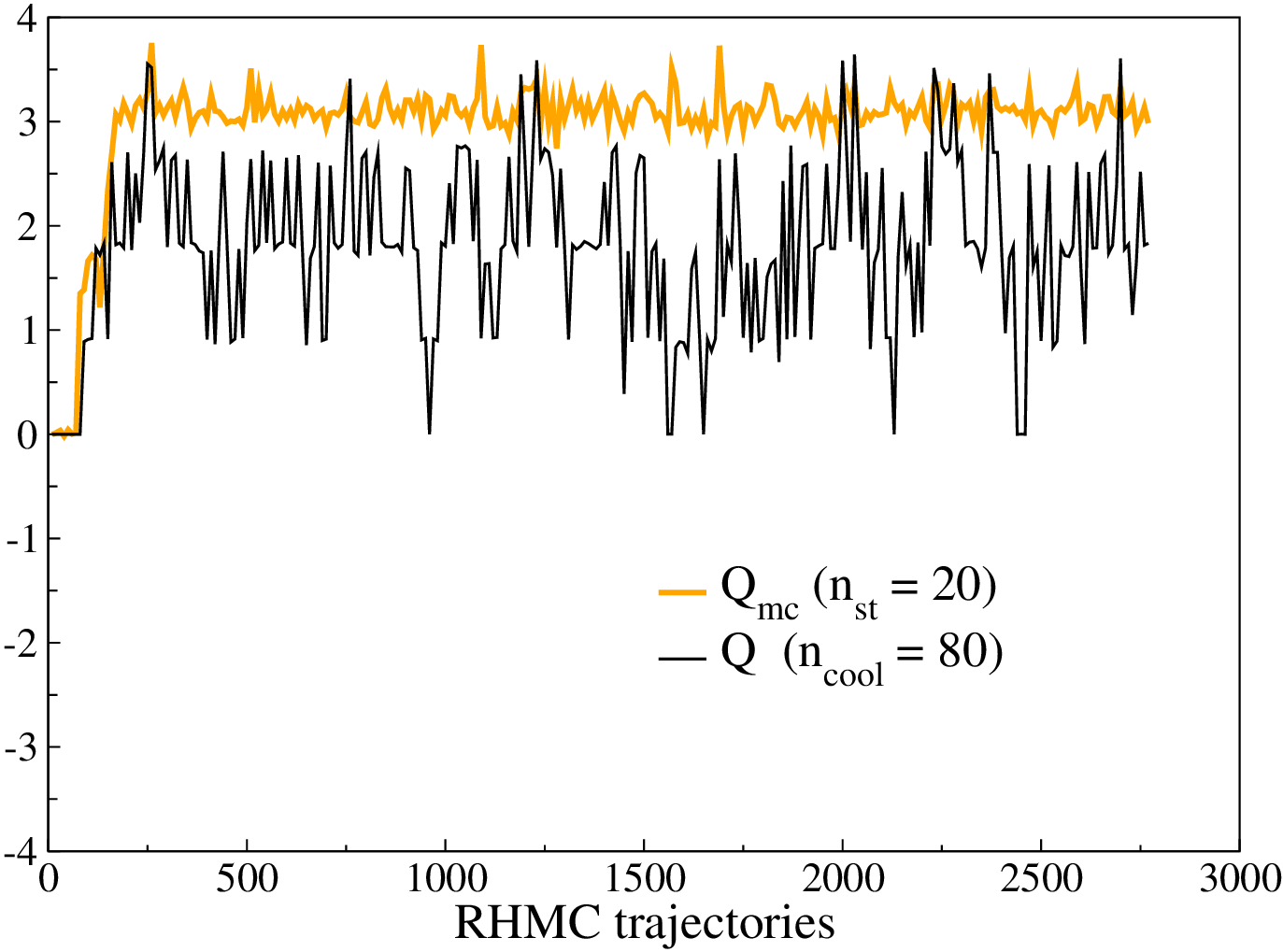}
\caption{Monte-Carlo history of the topological charges obtained
after 80 cooling steps and after 20 stout-smearing steps (with 
$\rho_{st} = 0.1$), for the run on the 
$32^3 \times 8$ lattice at $\beta = 4.14$, adopting 
the quadratic bias potential (see Eq.(\ref{potquad})) 
illustrated in Fig.~\ref{fig:potenziali}.
}
\label{fig:storiasbagliata}
\end{figure}

It is interesting to notice, looking at Fig.~\ref{fig:storiagiusta}, that the
correlation between the two charges, $Q$ and $Q_{mc}$ is visibly good; 
from a quantitative point of view one obtains, using the definition
in Eq.~(\ref{eq:correlation}), a good correlation around 0.86, meaning that the
bias on $Q_{mc}$ is also a good bias for $Q$. {For comparison, the
corrlation between $Q$ and $Q_{ft}$ before any stout smearing step is quite low
and around 0.08, while topological charges obtained after prolongated cooling
show of course larger correlation, for instance that between 60 and 80 cooling
steps is 0.97.}

The final estimate obtained for the susceptibility from this run is $a^4 \chi =
\langle Q^2 \rangle/V_4 = (6.1 \pm 1.1) \times 10^{-8}$, where the error has been
estimated after a binned jackknife analysis using Eq.~(\ref{reweight2}). Notice
that, after reweighting, the relative contribution to $\langle Q^2 \rangle$
from the $Q = 3$ sector turns out to be well below $0.01\, $, so that the cut
at $Q^{max}$, which does not {enhance} the
contribution from sectors with larger values of $Q$, is totally irrelevant.

In the following we will show results from several runs performed 
for different lattice volumes and at 
different values of the lattice spacing. In all cases we have adopted
a bias potential of the form showed in Eq.~(\ref{potrel}), fixing
the coefficient by reasonable guesses followed by some preliminary short
runs. For instance, changing the lattice volume at fixed
temperature and lattice spacing, $\langle Q^2 \rangle$ is expected
to scale proportionally to $V_4$, so that a good
starting guess is to rescale $B^2$ proportionally to $1/V_4$.

\section{Numerical results}
\label{results}

In this section, after discussing the effectiveness of our numerical 
approach, as compared to standard Monte-Carlo simulations, we will
illustrate results obtained for the topological susceptibility 
and for the $b_2$ coefficient, and discuss the relevance of finite size
and finite discretization effects.

\begin{figure}[t!]
\centering
\includegraphics*[width=0.65\columnwidth, clip]{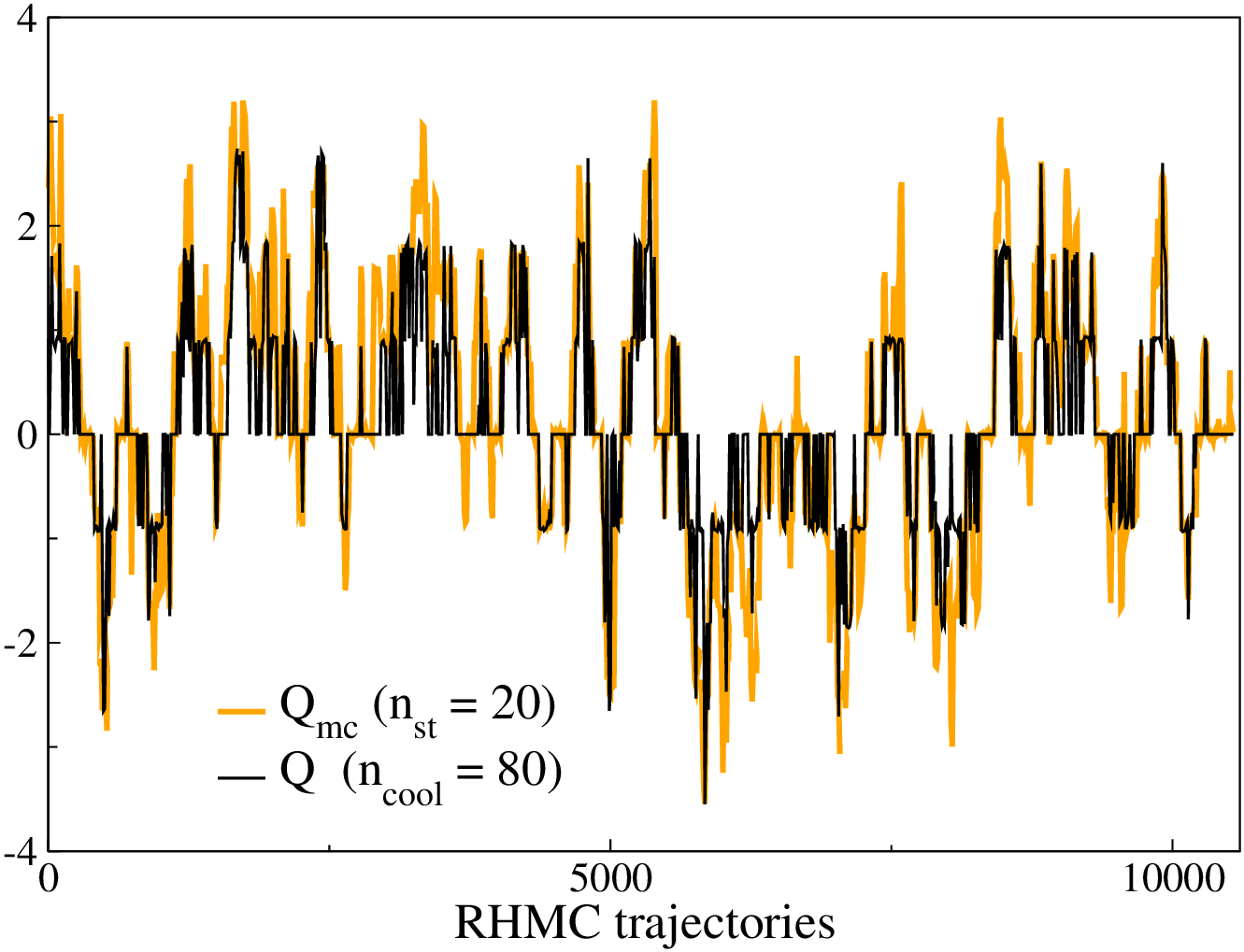}
\caption{Monte-Carlo history of the topological charges obtained
after 80 cooling steps and after 20 stout-smearing steps (with 
$\rho_{st} = 0.1$), for the run on the 
$32^3 \times 8$ lattice at $\beta = 4.14$, adopting a  
bias potential as in Eq.(\ref{potrel}), with $B = 6$ and $C = 2$,
and illustrated in Fig.~\ref{fig:potenziali}.
}
\label{fig:storiagiusta}
\end{figure}

\begin{figure}[t!]
\centering
\includegraphics*[width=0.65\columnwidth, clip]{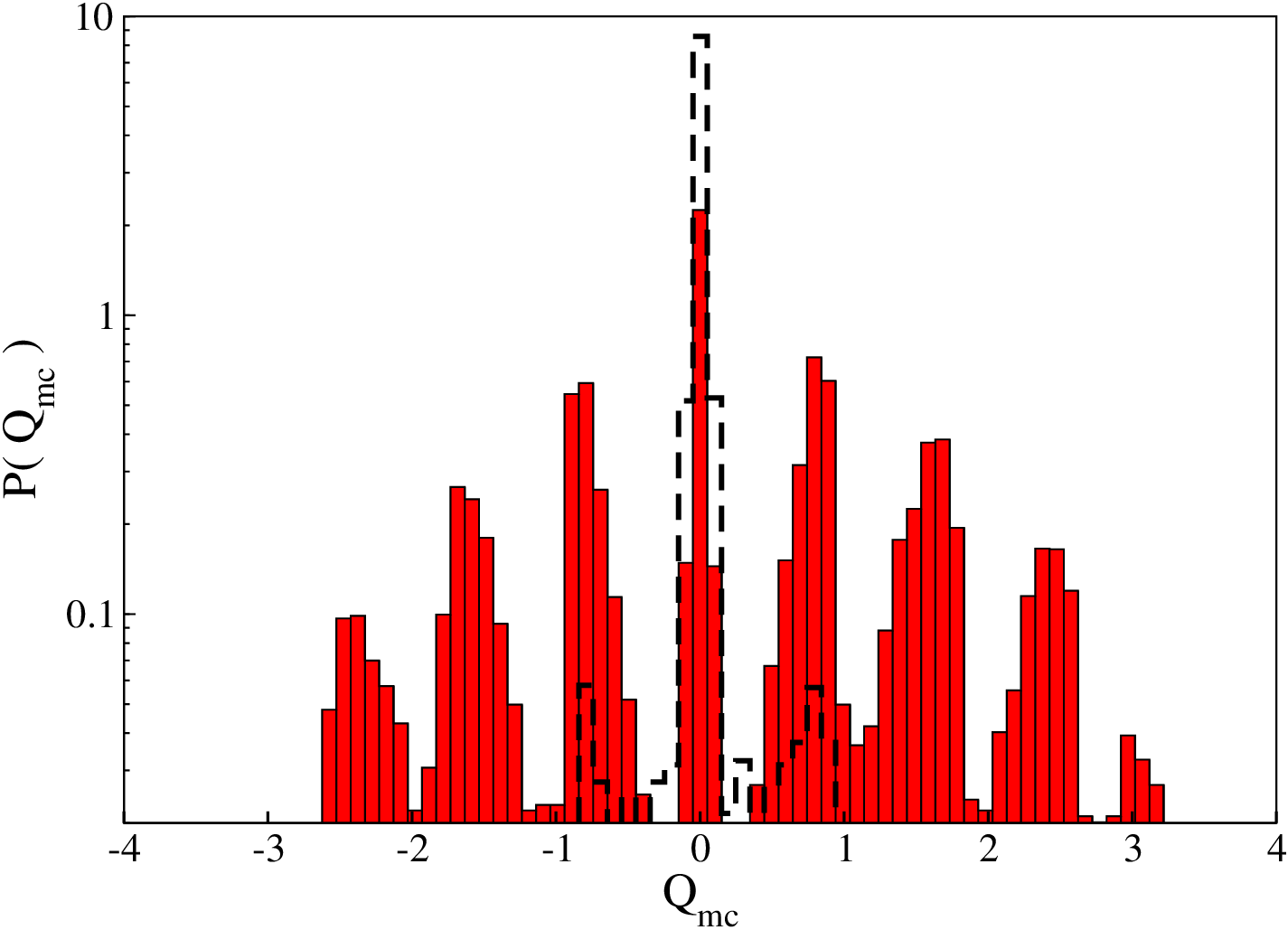}
\caption{Probability distribution of the topological charge $Q_{mc}$, measured
after 20 stout smearing steps with $\rho_{st} = 0.1$, for the same run
parameters and the same bias potential
relative to Fig.~\ref{fig:storiagiusta}. 
{As a reference, we also show (dashed line) the probability distribution
obtained with the same bare parameters at zero bias potential,
which has been already reported in Fig.~\ref{fig:histostout_4.1403208standard}.}
}
\label{fig:histostout_4.1403208stout}
\end{figure}

\subsection{Efficiency of the method and gain over the standard approach}

In the exploratory simulation that we have discussed in some detail 
in the previous section, the
relative error has decreased from 39\%, with the standard approach, to about
18\% when using the bias potential in Eq.~(\ref{potrel}), maintaining more or
less the same number of RHMC trajectories. That is as if we have gained roughly
a factor 4 in statistics, which however, when considering the 60\% overhead
required to integrate the equations of motion for $V(Q_{mc})$ with 20 stout
smaring steps, reduces to about a factor 2.5 in terms of computational effort
gain.

Of course, one should consider the additional moderate effort spent in the
preliminary small runs required to find a reasonable bias potential.  However,
the gain is expected to grow rapidly as the value of $\langle Q^2 \rangle$ one
has to determine decreases further: that could happen either by increasing the
temperature further or, according to the indications for a strong cut-off
dependence of $\chi$~\cite{Petreczky:2016vrs, Borsanyi:2016ksw}, by decreasing
the lattice spacing.

Therefore, let us consider a run performed on a $48^3 \times 16$ lattice at
$\beta = 4.592$, corresponding to $a \simeq 0.0286$~fm and the same temperature
$T \simeq 430$ MeV considered in the previous example. In this case we do not
have a simulation adopting the standard approach to compare with, since in that
case no topological fluctuation at all {is expected} in a few days run.
Instead, in Fig.~\ref{fig:pazzgain}, we show the Monte-Carlo (MC) history of the topological
charge over about $4.5 \times 10^4$ trajectories performed with a bias
potential as in Eq.~(\ref{potrel}), with $B =  11$ and $C = 2$. In this case,
the final result that we obtain after reweighting is $\langle Q^2 \rangle =
2.1(7) \times 10^{-4}$. 

From that we get a rough estimate of the computational effort that would have
been required to obtain the same statistical accuracy, which is around 30\%,
with the standard approach.  One should observe roughly $O(10)$ fluctuations
away from $Q = 0$: given the value of $\langle Q^2 \rangle$, that would require
around $5 \times 10^4$ independent draws of $Q$ when using the standard run.
We do not know what the autocorrelation time for $Q$ in the standard run would
have been, however we can assume it is at least of the same order of magnitude
as that observed in Fig.~\ref{fig:pazzgain}, which is $O(10^3)$.  Therefore,
one estimates the number of RHMC trajectories that would have been necessary
using the standard approach to be roughly between $10^7$ and $10^8$, implying a
gain in computational effort of $O(10^3)$.

\begin{figure}[t!]
\centering
\includegraphics*[width=0.65\columnwidth, clip]{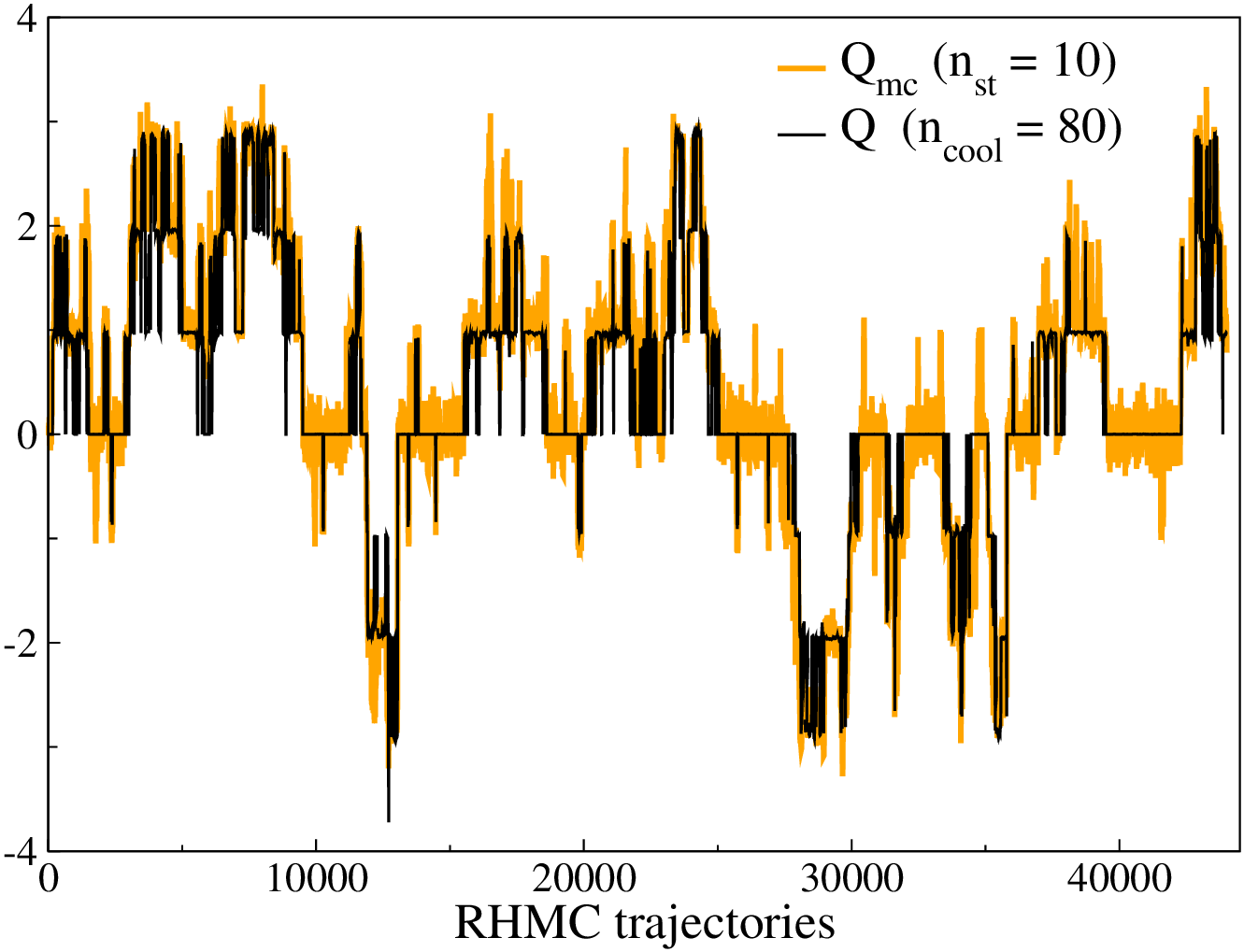}
\caption{
Monte-Carlo history of the topological charges obtained
after 80 cooling steps and after 10 stout-smearing steps (with 
$\rho_{st} = 0.1$), for the run on the 
$48^3 \times 16$ lattice at $\beta = 4.592$, adopting a 
bias potential as in Eq.(\ref{potrel}), with $B = 11$ and $C = 2$.
}
\label{fig:pazzgain}
\end{figure}

\subsection{Results for $\chi$ and $b_2$}

We now illustrate the numerical results
obtained for the topological susceptibility and for the $b_2$ coefficient at
two different temperatures and several lattice spacings, as illustrated in
Table~\ref{tab:bareparam}, with the aim of obtaining a continuum extrapolation.
Given the small values of $\chi$, one would also like to exclude {that}
the limited spatial volume available induces significant distortions in the
distribution of topological charge: therefore, at least for the lower
temperature, we have also explored, for every lattice spacing, different
spatial volumes, in order to exclude the possibile presence of finite size
effects.

\begin{figure}[t!]
\centering
\includegraphics[width=0.65\columnwidth, clip]{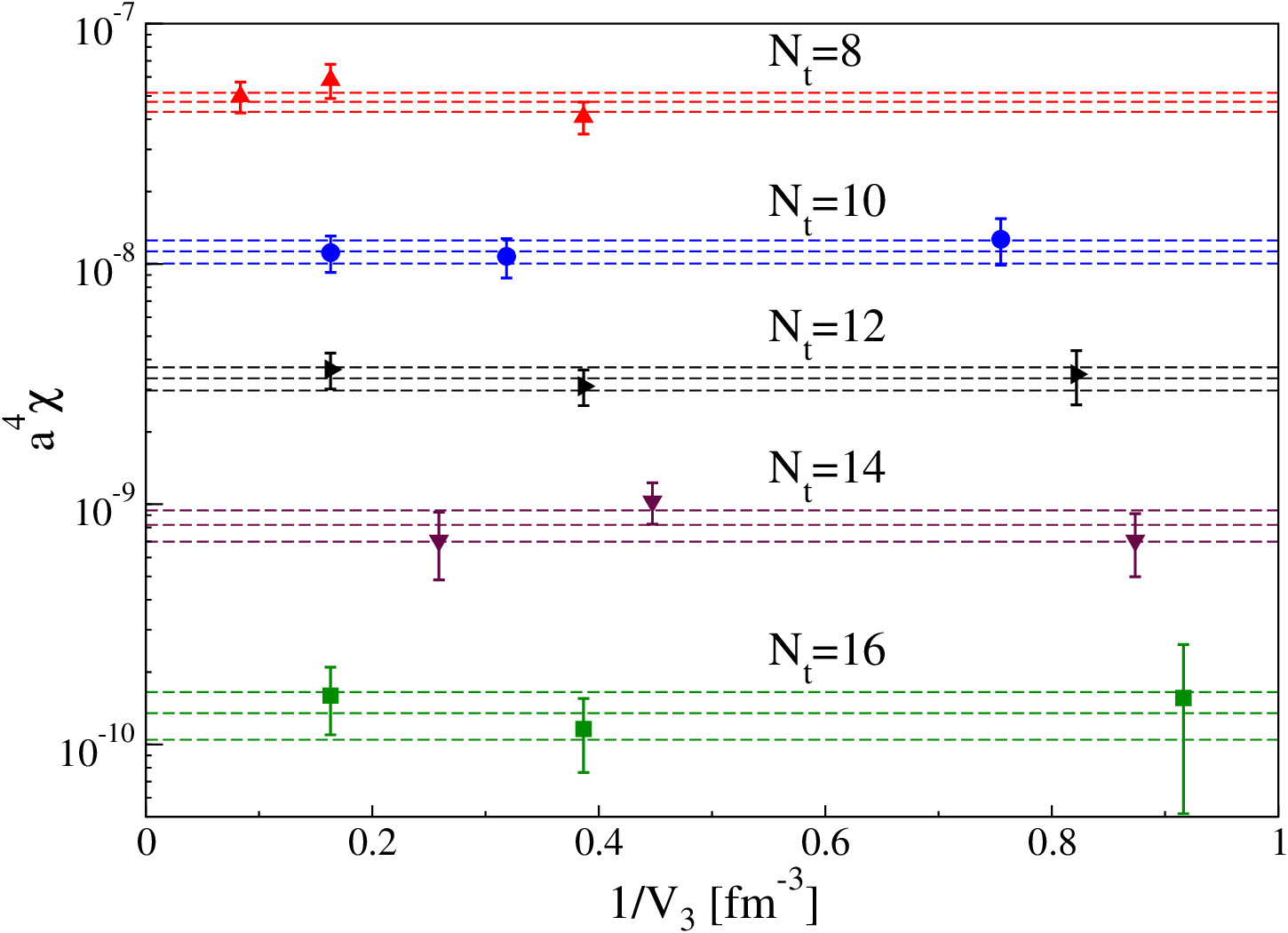}
\caption{Topological susceptibility in lattice spacing units, as a function
of the inverse spatial volume, for different values of $N_t$ at a fixed
value of the temperature, $T = 1 / (N_t a) \simeq 430$ MeV. The volume 
dependence is not significant and the horizontal bands are the result of 
a fit to a constant value.
}
\label{finitevolume430}
\end{figure}

\begin{figure}[t!]
\centering
\includegraphics[width=0.65\columnwidth, clip]{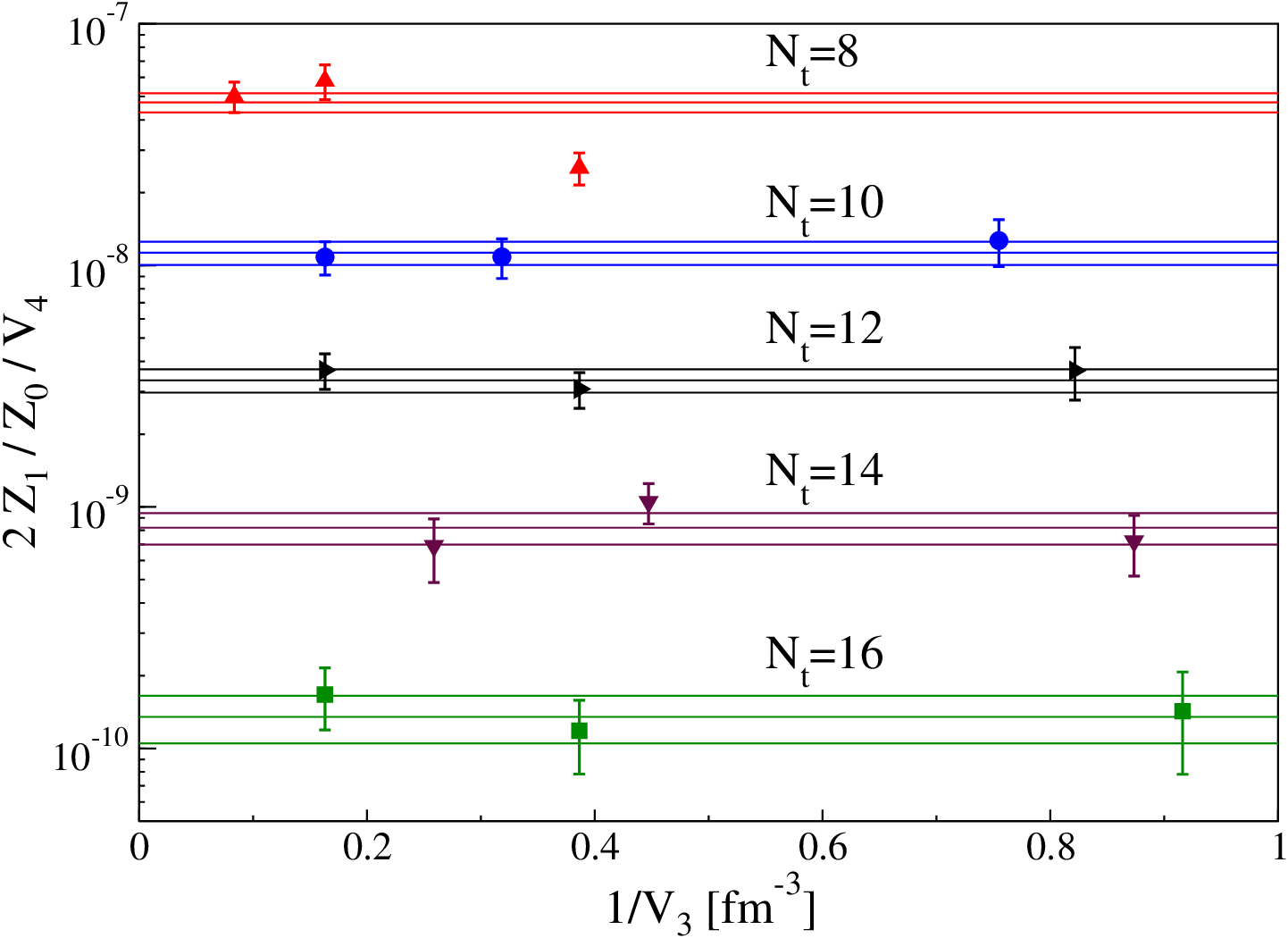}
\caption{{Ratios of the partition functions 
with $|Q| = 1$ to that with $|Q| = 0$, divided by the four volume $V_4$, for
the same lattice spacings and volumes shown 
in Fig.~\ref{finitevolume430}, and compared
with the infinite volume estrapolation of $a \chi^4$ 
(horizontal bands).}
}
\label{fig:2Z1suZ0}
\end{figure}

In Fig.~\ref{finitevolume430} we report the whole collection of results
obtained for $T \simeq 430$ MeV, where results  for $a^4 \chi$ are reported as
a function of the inverse spatial volume for the different lattice spacings
(see Table~\ref{tab:bareparam}), which appear in increasing order starting from
the bottom.  Finite size effects appear to be not significant, and the
horizontal bands represent our infinite volume estimates for $a^4 \chi$.

{It is interesting to compare the results obtained for 
$a^4 \chi$ with those that one would obtain by just sampling
the topological sectors with $|Q| = 0,1$: in this case the estimate
of $a^4 \chi$ would be given by $(2 Z_1 / Z_0)/V_4$, where 
$V_4$ is the four volume and $Z_Q$ 
is the partition function restricted to topological sector $Q$.
In general, the ratio $Z_{Q_1}/Z_{Q_2}$ 
can be simply obtained from our simulations by
taking the ratios between the  
reweighted averages for the occurences of $Q_1$ and $Q_2$.
In Fig.~\ref{fig:2Z1suZ0} we report 
$(2 Z_1 / Z_0)/V_4$ computed for the same lattice spacings
and volumes reported in Fig.~\ref{finitevolume430}: it is clear
that in fact, at least for the explored 
physical volumes, the two lowest topological sectors 
$|Q| = 0,1$ contain practically the whole information
relevant to the computation of the $a \chi^4$, whose
infinite volume extrapolation, already reported in 
Fig.~\ref{finitevolume430}, 
is reported again for a better comparison (horizontal bands).}

However, it is important to stress that our simulations
contain much more information than just 
$Z_1/Z_0$, indeed we are able to estimate
the ratios $Z_{Q_1}/Z_{Q_2}$ with good accuracy also for higher
values of $Q_1$ and $Q_2$. In Fig.~\ref{fig:2Z2suZ0} 
we report, as an example, the ratio $Z_2/Z_0$, normalized
by $V_4^2$ and measured for the different
lattice spacings and volumes. 
The reason of the normalization is to elucidate an important
piece of information contained in these data: $Z_2/Z_0$ scales
proportionally to $V_4^2$, as expected for the occurrence 
of two independent and non-interacting topological objects 
in the same volume, i.e.~for the Poissonian distribution of 
the topological charge predicted by DIGA.
A similar
behaviour, i.e.~a scaling with the appropriate
power of the volume predicted by DIGA, is observed also for $Z_3/Z_0$.

\begin{figure}[t!!]
\centering
\includegraphics[width=0.65\columnwidth, clip]{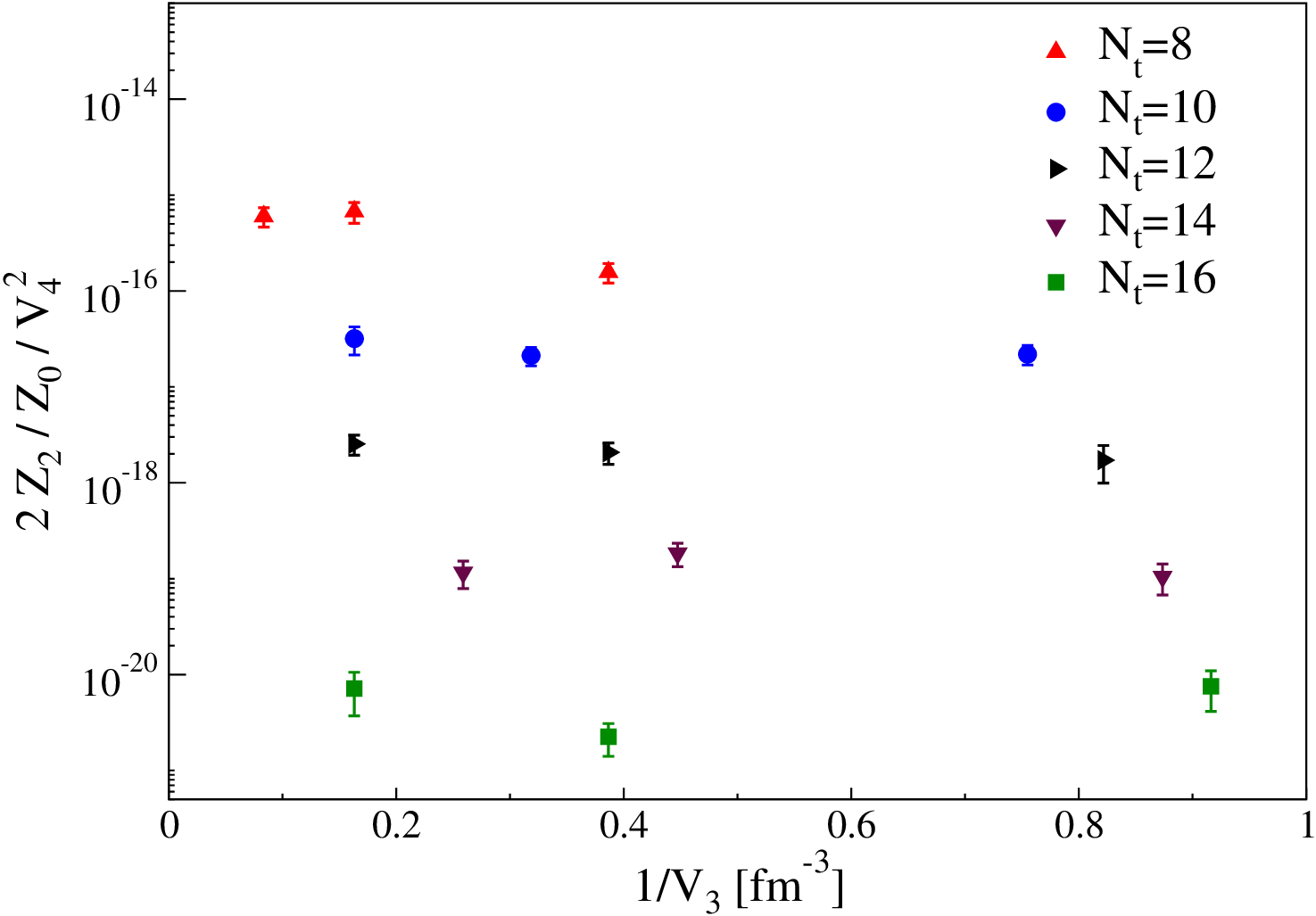}
\caption{{Ratios of the partition functions 
with $|Q| = 2$ to that with $|Q| = 0$, divided by $V_4^2$, for
the same lattice spacings and volumes shown 
in Fig.~\ref{finitevolume430}.}
}
\label{fig:2Z2suZ0}
\end{figure}

The same estimates, expressed in physical units for $\chi^{1/4}$, are reported
in Fig.~\ref{continuumlimit430} as a function of $a^2$; in the same figure,
continuum extrapolations at the same temperature from Ref.~\cite{axion_1} and
Ref.~\cite{Borsanyi:2016ksw} are also reported.  As one can easily appreciate,
finite cut-off effects are huge, with $\chi$ decreasing by a factor 40 when
moving from $a \simeq 0.0576$ fm to $a \simeq 0.0286$ fm: that explains the
discrepancy observed among previous literature results.  Continuum
extrapolation is possible with our present data: we require to include $O(a^4)$
corrections when considering all explored lattice spacings, and just $O(a^2)$
for the three smallest ones: results are consistent and we obtain {$\chi^{1/4}
= (3.2 \pm 3.3)$ MeV} ($\tilde\chi^2 = 7.6/2$) in the first case and
{$\chi^{1/4} = (5.3 \pm 3.1)$ MeV} ($\tilde\chi^2 = 2.8/1$) in the second case,
both compatible, even if within large error bars, with the results reported in
Ref.~\cite{Borsanyi:2016ksw}. A similar agreement is observed for results
obtained at $T = 570$ MeV, which are reported in Fig.~\ref{continuumlimit570}:
in this case we obtain {$\chi^{1/4} = (6.6 \pm 4.8)$ MeV} ($\tilde\chi^2 =
1.3/1$) when considering $O(a^4)$ corrections for all lattice spacings and
{$\chi^{1/4} = (13 \pm 3)$ MeV} ($\tilde\chi^2 = 2.4/1$) when considering just
$O(a^2)$ corrections for the three smallest lattice spacings.  

In almost all the cases the values of $\tilde{\chi}^2$ are somehow
large, however (because of the small number of degrees of freedom) these
numbers are not incompatible with the hypothesis that $\chi^{1/4}(a)$ is linear
in $a^2$ for the three smallest 
values of the lattice spacing. On the other hand
the large $\tilde{\chi}^2$ value obtained for $T=430$\,MeV when using all data
points, and the only marginal agreement between the results of the two fits
performed at $T=530$\,MeV, are indications that the continuum limit systematics
are still not completely under control.
A fair account of our final results, based on the fits with
lowest chisquared but including such systematics,
would be $(3 \pm 3 \pm 2)$~MeV for $T=430$\,MeV and 
$(7 \pm 5 \pm 6)$~MeV for $T=530$\,MeV.
Moreover, the almost $100\%$ relative
errors of our continuum extrapolations prevent a reliable estimate of the
power law coefficient in Eq.~(\ref{eq:diga_chi}).  Let us however stress once
more that our main purpose in this paper was not to provide a precise
determination of the temperature dependence of $\chi(T)$, but to show that the
obstacle represented by the suppression of the $Q\neq 0$ sectors at high
temperature can be overcome in a model independent way. This is obviously not
the end of the story since we still have other obstacles and, as we discuss in
the next section, smaller lattice spacings or a way to reduce lattice artifacts
are needed for the future.

\begin{figure}[t!]
\centering
\includegraphics[width=0.65\columnwidth, clip]{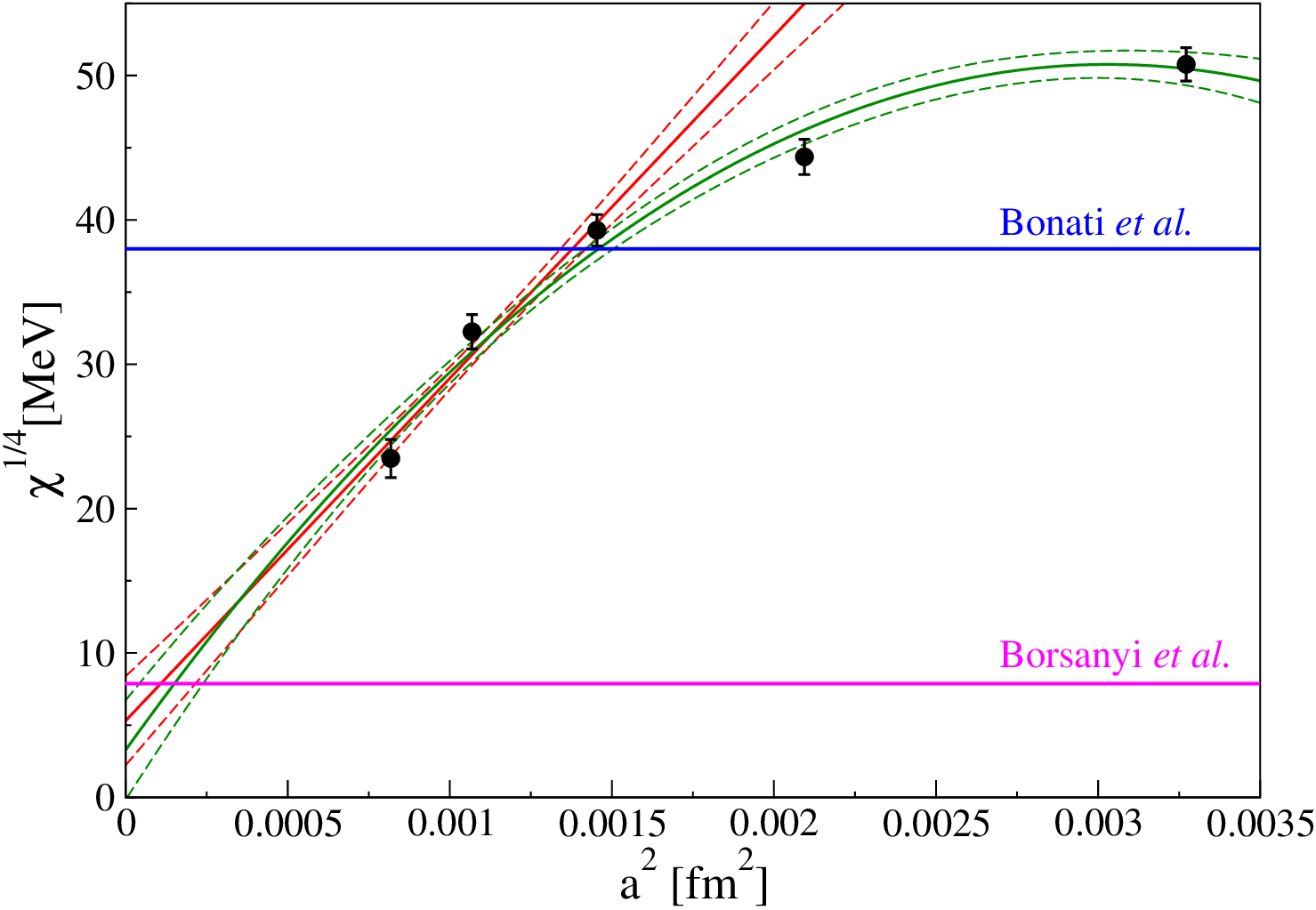}
\caption{Fourth root of the topological susceptibility, as a function of 
$a^2$, for $T \simeq 430$ MeV. To two bands represent the result
of two different continuum extrapolations, taking into account
respectively $O(a^2)$ or $O(a^4)$ corrections.
The two horizontal lines are the continuum extrapolations reported 
respectively in Refs.~\cite{axion_1} and \cite{Borsanyi:2015cka}
at the same temperature. 
}
\label{continuumlimit430}
\end{figure}

Despite the huge cut-off effects observed for $\chi$, the values obtained for
the $b_2$ coefficient defined in Eq.~(\ref{eq:b2}), which are reported in
Fig.~\ref{continuumlimitb2430} for $T = 430$ MeV, are practically independent
of $a$ within errors, and always compatible with the prediction from DIGA, $b_2
= - 1/12$. {A fit to all lattice spacings at this temperature
assuming $O(a^2)$ corrections yields $- 12 b_2 = 1.0006(10)$ 
($\tilde\chi^2 = 0.75/3$), while 
$- 12 b_2 = 0.9997(3)$ 
($\tilde\chi^2 = 1.6/4$) is obtained taking a fit to a constant
function.}
These results for $b_2$ are consistent with 
the assumption of a dilute gas of independent
topological objects, however let us stress
that we could not have expected anything different from 
that, given the fact that, on the explored volumes, 
the topological sectors with $|Q| = 0,1$ are largely
dominant (see Fig.~\ref{fig:2Z1suZ0}) and moreover $Z_1 / Z_0 \ll 1$:
that alone yields inevitably
to $b_2 \simeq -1/12$ and would happen even for $T = 0$ if one
takes the volume small enough.
Therefore, a more substantial support to the validity of 
DIGA, which is assumed in 
Ref.~\cite{Borsanyi:2016ksw}, is given 
by the analysis of the volume scaling 
of the ratios $Z_{Q}/Z_0$ that we have 
discussed above: that guarantees that, even 
when making the volume large enough that 
$|Q| = 1$ does not dominate any more, $b_2$ will 
remain consistent with DIGA and the susceptibility will
maintain the value already determined on the
smaller, accessible volumes.

\section{Discussion}
\label{discussion}

The determination of the topological susceptibility in the high temperature
regime has to face various numerical challenges: topological fluctuations
become very rare and difficult to sample, UV
cut-off effects can be quite large because of the 
{non-exact lattice chiral symmetry} of the adopted fermion
discretization and, finally, the freezing of topological modes leads to a
critical slowing down when one gets too close to the continuum limit.  These
difficulties are at the basis of the discrepancies found among recent lattice
determinations adopting different approximations or assumptions.

In this study, we have shown how rare topological events can be effectively
sampled, in a controlled way, by inserting a bias $Q$-dependent potential in
the probability distribution of gauge configurations, which is then reweighted
away {in the analysis}. That permits to gain orders of magnitude in terms of
computational effort, in situations in which $\langle Q^2 \rangle \ll 1$.
{That has given access not only to the topological susceptibility and to $b_2$,
but also to quantities like $Z_Q/Z_0$ with $|Q| > 1$, which are extremely small
on the accessible lattice volumes and would have been completely unaccessible
otherwise: a careful check of the volume scaling has revealed that finite size
effects are not relevant and has given substantial support to the validity of
the  DIGA assumption in the explored temperature regime.}

\begin{figure}[t!]
\centering
\includegraphics[width=0.65\columnwidth, clip]{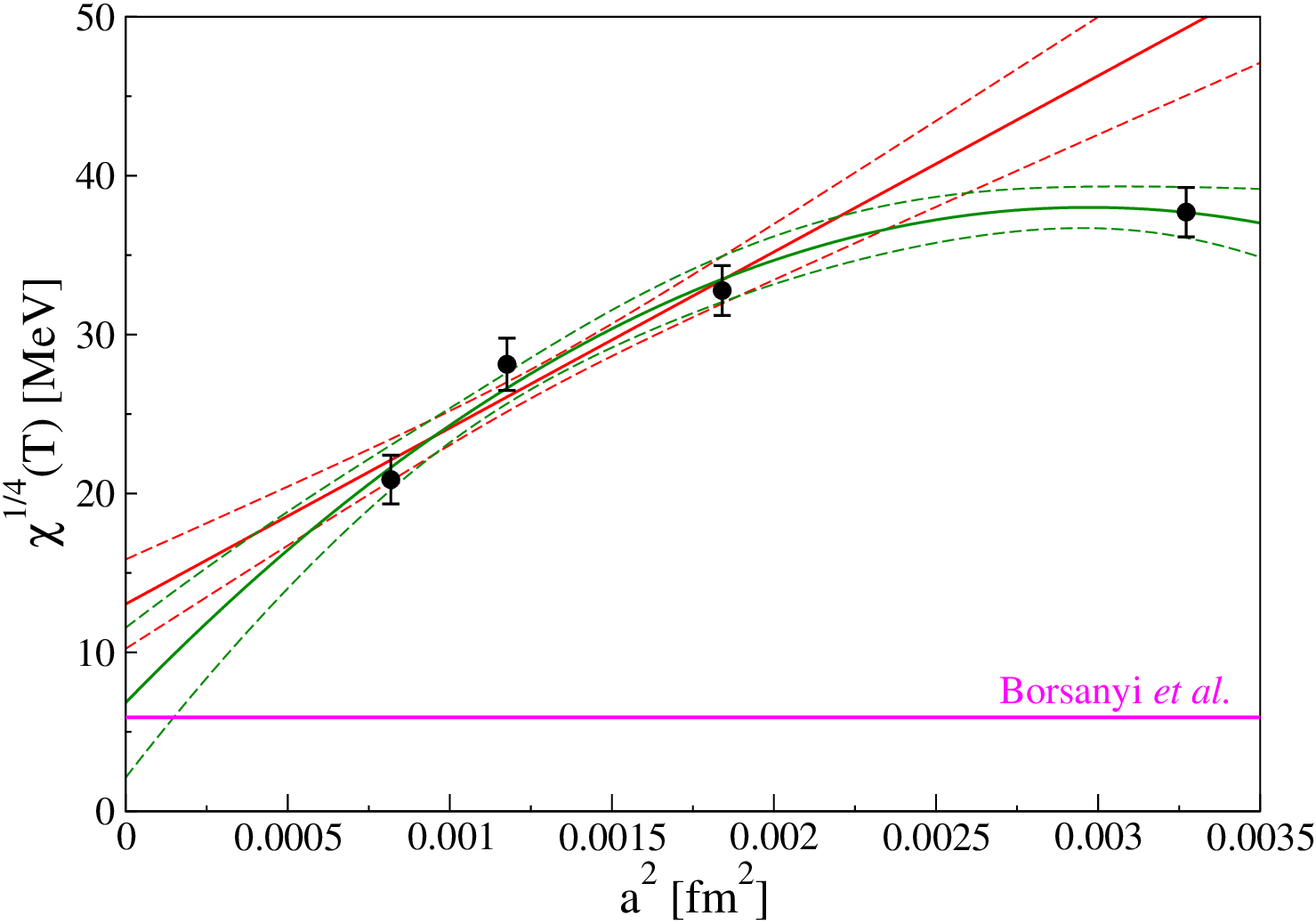}
\caption{Fourth root of the topological susceptibility, as a function of 
$a^2$, for $T \simeq 570$ MeV. To two bands represent the result
of two different continuum extrapolations, taking into account
respectively $O(a^2)$ or $O(a^4)$ corrections.
The horizontal lines is the continuum extrapolation reported 
in Ref.~\cite{Borsanyi:2015cka}
at the same temperature. 
}
\label{continuumlimit570}
\end{figure}

Nevertheless, as we have seen, the large UV cut-off effects still represent a
problem: $\chi$ drops by 3-4 orders of magnitude when moving from $a \sim 0.06$
fm towards $a = 0$, meaning that, despite the huge improvement induced by the
bias potential, the final continuum extrapolation still has considerable error
bars.  In particular, the present accuracy obtained at the two explored
temperatures, with relative errors not far from 100\%, does not permit us to
extract a reliable estimate for the power law coefficient in
Eq.~(\ref{eq:diga_chi}).

In Ref.~\cite{Borsanyi:2016ksw}, UV effects have been suppressed by an ad hoc 
procedure which reweights gauge configurations with $Q \neq 0$ 
by forcing the $Q$ lowest eigenvalues of the discretized Dirac operator
associated with them to be zero. This procedure becomes exact as
one approaches the continuum limit, where the Dirac operator
indeed develops exact zero modes, however it induces non-local
modifications in the discretized theory at finite lattice spacing.
For that reason, an independent way of obtaining continuum
extrapolations of $\chi$ with a good statistical 
accuracy would be welcome.

\begin{figure}[t!]
\centering
\includegraphics[width=0.65\columnwidth, clip]{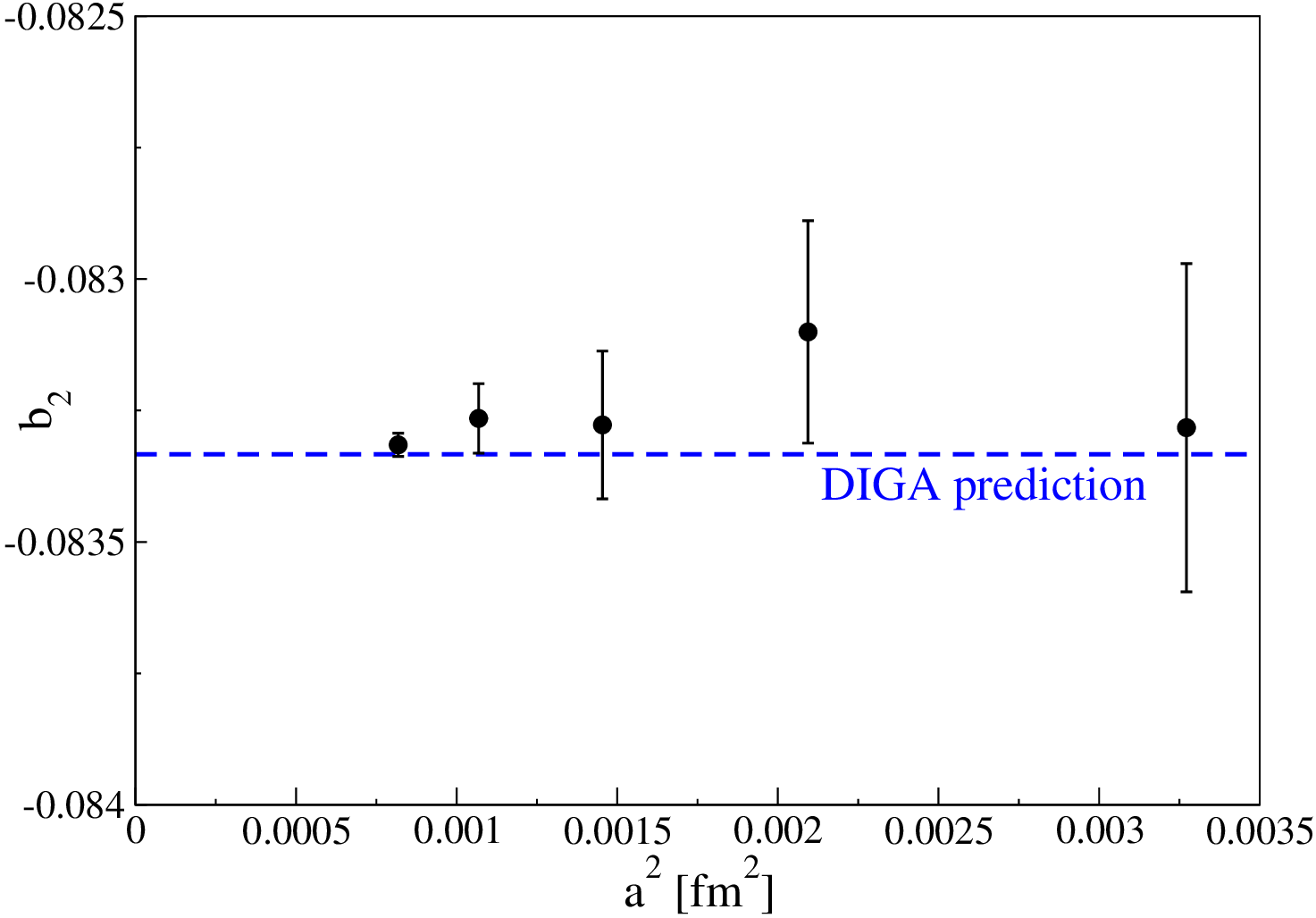}
\caption{Numerical results obtained for the coefficient $b_2$,
as a function of $a^2$, for $T \simeq 430$~MeV. 
The horizontal line represents the prediction of the 
dilute instanton gas approximation (DIGA).
}
\label{continuumlimitb2430}
\end{figure}

An obvious strategy is to push our efforts forward by exploring smaller lattice
spacings. However, that will require to face problems related to critical
slowing down, since at the smallest explored lattice spacing the
autocorrelation length is already larger than $O(10^3)$ unit time 
RHMC trajectories. A
natural solution could be represented by metadynamics~\cite{LaioParr,
LaioGervasio, Laio:2015era}, where the bias potential is dynamically tuned so
as to enhance the tunneling between different topological sectors.

A different approach could be represented by a modification of the observable
used to determine the topological susceptibility.  
{Present results are based on a gluonic definition of 
the topological charge relying 
on the smoothing of gauge configurations. 
The adoption of a fermionic definition of the topological charge
could be a much better choice: the point here is not about 
the solid theoretical basis of this definition, but rather
about the practical benefits that one could 
achieve if the same discretization of the 
Dirac operator is adopted both for the MC simulation
and for the determination of the topological content.} 
In particular, recent
studies performed at zero temperature~\cite{Alexandrou:2017bzk} have shown that
definitions of $\chi$ based on spectral projectors~\cite{Giusti:2008vb} lead to
finite cut-off effects which are significantly reduced with respect to those
based on standard gluonic observables. We plan to consider both strategies in
the near future.

\acknowledgments
We thank A.~Athenodorou for useful discussions.
We acknowledge PRACE for awarding us access to resource MARCONI 
based in Italy at
CINECA, under project Pra13-3331 - AXTRO.  
FN acknowledges financial support from the INFN HPC\_HTC project.

\end{document}